\documentstyle[twocolumn,prb,aps,epsf,epsfig,amsfonts]{revtex}
\begin{document}

\title{A study of the tunnelling-charging
Hamiltonian of a Cooper pair pump}

\author{M. Aunola}

\address{Dept. of Physics, University of Jyv\"askyl\"a,
P.O. Box 35 (Y5), FIN-40351 Jyv\"askyl\"a, Finland\\
Email: Matias.Aunola@phys.jyu.fi}

\date{\today}
\maketitle

\begin{abstract}
General properties of the tunnelling-charging Hamiltonian of a 
Cooper pair pump are examined with emphasis on the symmetries of 
the model. An efficient block-diagonalisation scheme and a
compatible Fourier expansion of the eigenstates is constructed
and applied in order to gather information on important 
observables. Systematics of the adiabatic pumping
with respect to all of the model parameters  are obtained
and the link to the  geometrical Berry's phase is identified.
\end{abstract}

\vskip0.5truecm {\noindent Adiabatic transport of single electrons in
arrays of small metallic tunnel junctions has been widely studied in
recent years.\cite{pot92,kel96}} In the Coulomb blockade regime
phase-shifted gate voltages have been used to induce a dc current
$I=-nef$, where $n$ is the number of carried electrons and $f$ is the
gating frequency. Normal-state pumps transporting single electrons
have reached accuracy that can be considered for metrological
applications.\cite{kel96} Pumping of Cooper pairs has gained
interest due to new ideas in quantum measuring and
computing.\cite{ave98,shn98}

A quantitative theory of pumping Cooper pairs in gated arrays of
Josephson junctions when the environment has negligible impedance has
been presented.\cite{pek99,aun00} The leading order pumped current is
$I\approx-2ef[1-a_{N}(\varepsilon_{\rm J} )^{N-2}\cos\phi]$, where
$a_N$ is a constant and $\varepsilon_{\rm j}:=E_{\rm J}/ E_{\rm C}$ is
the coupling strength. Here $E_{\rm J}$ and $E_{\rm C}:=(2e)^2/2C$
are the Josephson
coupling energy and the charging energy, respectively. According to
recent calculations the $\cos\phi$-dependent inaccuracy should be
experimentally observable at least in a certain frequency
range.\cite{pek00} In this article the model for adiabatic pumping from
Refs.~\ref{bib:pek} and \ref{bib:aun} is examined in a thorough
manner, concentrating on the symmetries and general features of the
simplified system, thus extending the treatment to strong couplings and
long arrays. The overall behaviour of the pumped current is
explained and systematised, but not rigorously proven.

A schematic view of a Cooper pair pump and
the ideal operation of the gate voltages $V_{{\rm g},k}$ are shown in
Fig.~\ref{fig:cppump}. On any of the $N$ legs of a cycle at most
two gate voltages are changed. The tunnelling-charging Hamiltonian
\begin{equation}
H=H_{\rm C}+H_{\rm J}\label{simphami},
\end{equation}
neglects the quasiparticle tunnelling and other degrees of
freedom. The full set of  model parameters are  $\varepsilon_{\rm J}$,
total phase difference over the array
$\phi$, the relative junction capacitances $\vec c$, where
$c_k:=C_k/C$ and $\sum_{k=1}^NC_k^{-1}=N/C$, and the
(normalised) gate charges $\vec q:=\{q_1,\ldots,q_{N-1}\}$, where
$q_k:=-C_{{\rm g},k} V_{{\rm g},k}/2e$. For homogeneous arrays 
$c_k:=1$ and the inhomogeneity is quantified by $X_{\rm inh}=
[\sum_k (c_k^{-1}-1)^2/N]^{1/2}$.
The model Hamiltonian is diagonal with respect to $\phi$,
which is assumed to be fixed, but could be controlled by an
external bias voltage  according to $d\phi/dt=-2eV/\hbar$ and
is subject to any  voltage fluctuations.\cite{pek00,ing92}
The conjugate variable $\hat M$, the average number of tunnelled
Cooper pairs, is undetermined in the present model.

\vbox{
\begin{figure}
\begin{center}
      \mbox
      {\epsfig{file=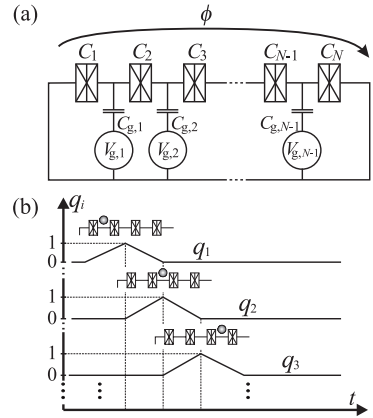,height=55truemm}
      }
\end{center}
\caption{(a) A schematic drawing of a gated Josephson array of
$N(\ge3)$ junctions. The $C_k$ and $C_{{\rm g},k}$ are the
capacitances  of the junctions and gates, respectively. 
b) The optimal operation of  the gate charges $\vec q$.
Each cycle carries approximately one Cooper pair through the array,
when $\varepsilon_{\rm J}\ll1$.
\label{fig:cppump}
}
\end{figure}
}

The matrix elements of the charging Hamiltonian $H_{\rm C}$ are given by 
the capacitive charging energy 
\begin{equation}
\langle \vec n\vert H_{\rm C}(\vec q)\vert \vec n\rangle_{\phi}=
E_{\rm C}\left[\sum_{k=1}^N\frac{v_k^2}{c_k}
-\frac1N\left(\sum_{k=1}^N\frac{v_k}{c_k}\right)^2\right],
\label{generchar0}
\end{equation}
where the number of Cooper pairs on each island is given 
by $\vec n$. The quantities $v_k$, $k=1,\ldots,N$, are a
solution of 
\begin{equation}
v_k-v_{k+1}=n_k-q_k. \label{eq:veeesi}
\end{equation}
Tunnelling of one Cooper pair through the $k$th
junction changes $\vert\vec n\rangle$ by $\vec\delta_k$,
where the non-zero components are (if
applicable) $(\vec \delta_k)_k=1$ and $(\vec \delta_k)_{k-1}=-1$.
The tunnelling Hamiltonian then reads
\begin{equation}
H_{\rm J}=-\sum_{\vec n,k=1}^{N}\frac{c_kE_{{\rm J}}}{2}(\vert
\vec n+\vec\delta_k\rangle\langle \vec n\vert e^{i\phi/N}+\,{\rm H.c.}\,).
\label{eq:tunnel}
\end{equation}
The supercurrent flowing through the array is determined by
the supercurrent operator
\begin{equation}
I_{\rm S}=(-2e/\hbar)(\partial H/\partial \phi),\label{averagesup}
\end{equation}
a G$\hat{\mathrm a}$teaux derivative\cite{rud73} of the
full Hamiltonian. By changing the gate voltages adiabatically along a
closed path $\Gamma$, a charge transfer
$Q_{\rm tot}:=Q_{\rm s}+Q_{\rm p}$ is induced.
The pumped charge, $Q_{\rm p}$, depends only on
the chosen path, while the charge due to direct supercurrent, $Q_{\rm s}$,
also depends on how the  gate voltages are operated.
If the system remains in a
adiabatically evolving state $\vert m\rangle$, the total transferred 
charge, $Q_{\rm tot}$, in units of $-2e$ reads\cite{pek99,aun00}
\begin{eqnarray}
\frac1{\hbar}\int_0^\tau
\frac{\partial E_m(t)}{\partial \phi}dt-2\oint_{\Gamma}
\sum_{l(\ne m)}{\rm Im}\left[\frac{\langle m\vert I_{\rm S}
\vert l\rangle\langle l\vert dm\rangle}{E_l-E_m}\right],
\label{eq:norpump}
\end{eqnarray}
where $\vert dm\rangle$ is the
change in $\vert m\rangle$ due to a differential change of the gate
voltages $d\vec q$. 

Intermediate states $\vert l\rangle$ and energy
denominators can be removed by
rewriting the off-diagonal matrix element
between stationary states as
\begin{equation}
\langle m\vert \partial H/\partial \phi\vert l\rangle=
(E_l-E_m)[\partial(\langle m\vert)/\partial\phi]\vert l\rangle,
\end{equation}
and using the identity $\sum_l\vert l\rangle\langle l\vert=\hat 1$.
The canonical representation $\hat M=-i\partial/\partial\phi$
connects the pumped charge to
the average number of tunnelled Cooper pairs by
\begin{equation}
Q_{\rm p}=2\oint_{\Gamma}{\rm Re}\left[\langle m\vert\hat M
\vert dm\rangle\right].
\label{eq:simppump}
\end{equation}
A single eigenstate for phase differences $\phi$ and $\phi+d\phi$
is required for each integration point. This expression
identifies $\hat M$ as the link between $Q_{\rm p}$ and
the geometrical Berry's phase,
$\gamma_m(\Gamma)=i\oint_{\Gamma} \langle m\vert dm\rangle$.\cite{ber84}
The connection was given without an explicit
identification in Ref.~\ref{bib:pek} and was
mentioned in Ref.~\ref{bib:fal}.
For $\varepsilon_{\rm J}\rightarrow 0$ only two charge
are of importance during each leg. From the wave function
$\vert m\rangle=[(1-a^2)^{1/2},\,ae^{i\phi/N}]^T$,
where $a:0\rightarrow 1$, one obtains $\langle m\vert\hat M
\vert dm\rangle=d(a^2)/2N$ and $Q_{\rm p}=1$ for a full cycle.

A reference state $\vert \vec n_0\rangle$ induces
a convenient labelling of the charge states.
Each state is denoted by integers $\{ y_k\}_{k=1}^N$,
$0\le Y_{\vec n}:=\sum_k y_k<N$, such that $\vec n=\vec n_0
+\sum_k y_k\vec \delta_k$. The numbers $y_k$ tabulate the
number and direction of tunnellings from $\vec n_0$ to $\vec n$.
The distance between charge states is defined by
$\tilde d(\vec n_1,\vec n_2):=\min\left(\sum_{k=1}^N\vert y_{k}^{(1)}
-y_{k}^{(2)}+l\vert:l\in{\mathbb Z}\right)$.
If $\tilde d(\vec n_1,\vec n_2)=1$
($=l$), then $\vert\vec n_1\rangle$
and $\vert\vec n_2\rangle$ are ($l$th) nearest neighbours.

A change of basis $U\{\vert \vec n\rangle\}=
\{\vert e^{i\phi Y_{\vec n}/N}\vec n\rangle\}$
yields a Hamiltonian matrix $\tilde H=UHU^{-1}$  with properties
$\tilde H(\phi+2\pi)=\tilde H(\phi)$ and $\tilde H(-\phi)=(
\tilde H(\phi))^*$.  Thus eigenstates and eigenvalues of
$\tilde H$ are periodic under $2\pi$, or state
labels change cyclically. Usually the former happens, so
the supercurrent in a stationary state is described by
a Fourier sine series
\begin{equation}
\langle I_{\rm S}\rangle_{(N,\varepsilon_{\rm J},\phi,\vec c,\vec q)}:=
\hbox{$\sum_{l=1}^{\infty}$}\alpha_l\sin(l\phi).\label{eq:superc}
\end{equation}
The ground state supercurrent behaves differently only
at the so-called resonance points,
where the ground state becomes  degenerate
for $\phi=\pi+2l\pi$, $l\in{\mathbb Z}$.
For homogeneous arrays resonance points are located at
$\vec q=\vec n\pm(1,\ldots,1)/N$, where $\vec n$ is arbitrary.
The corresponding poles in the Berry's phase
give raise to $Q_{\rm p}$, and $\varepsilon_{\rm J}$
determines which poles are important.

Due to symmetry of the representation $\tilde H$ the original
amplitudes in $2\pi$-periodic states are given by
\begin{equation}
a_{\vec n}^{\vec q,\phi}=\sum_{l=-\infty}^\infty
a_{\vec n,l}^{\vec q}e^{i\phi(l+Y_{\vec n}/N)},\label{eq:amplit}
\end{equation}
where real Fourier coefficients $a_{\vec n,l}^{\vec q}$ are
fixed by the gauge condition
$a_{\vec n'}^{\vec q,\phi}:=\vert a_{\vec n'}^{\vec q,\phi}\vert
e^{i\phi Y_{\vec n'}/N}$ for the charge state
$\vert \vec n'\rangle$.  
The averaged number of tunnelled Cooper pairs,
\begin{equation}
{\cal M}:=\hbox{$\sum_{\vec n,l}$}(l+Y_{\vec n}/N)(a_{\vec n,l}^{\vec q})^2,
\end{equation}
is unique up to a gauge-dependent integer.  The apparent contradiction
between sharp phase difference combined with sharp value of $\cal M$
is an artefact due to gauge-fixing. Stronger coupling increases
the variance of $\cal M$. Discontinuities 
in gauges certainly occur on any closed path encircling an odd number
of resonance points. A gauge is unstable near a discontinuity,
but everywhere away from resonance points many valid gauges exist.
Especially on the gating path depicted in Fig. 1, the 
dominant charge states on each leg give very stable gauges. 

The pumped charge,  $Q_{\rm p}$, can be evaluated using
a gauge-independent differential expression
\begin{eqnarray}
dQ_{\rm p}(\phi)
&=&\sum_{l'=0}^\infty\sum_{\vec n,l=-\infty}^\infty\left[
\frac{2(l+Y_{\vec n}/N)}{1+\delta_{l'0}}d(a_{\vec n,l}a_{\vec n,l+l'})
\right.\cr
&&\ +\left. l'(a_{\vec n,l}da_{\vec n,l+l'}-
a_{\vec n,l+l'}da_{\vec n,l})\right]\cos(l'\phi),
\label{eq:transchar}
\end{eqnarray}
where the $\phi$-independent average is simply $d{\cal M}$.
Due to the normalisation of $\vert m\rangle_{\phi}$,
the coefficients are orthonormal, i.e.
$\sum_{\vec n,l}a_{\vec n,l}a_{\vec n,l+l'}=\delta_{l'0}$,
which cancels the terms multiplying the full differential by $l'$.
Expression (\ref{eq:transchar}) indicates that the averaged charge transfer 
for a full cycle is is exactly $-2e$, regardless of the inhomogeneity of
the array or reasonable deformations of the gating path.

The tunnelling-charging Hamiltonian
can often be block diagonalised with the following
transformation. Let an orthonormal basis $\{\vert s\rangle\}$ span
the Hilbert space ${\cal H}$ and
the matrix elements of a Hamiltonian $H$
be $h_{ss'}:=\langle{s}\vert H\vert {s'}
\rangle$. Choose projection operators $\{P_i\}$
by $P_i=\sum_{k=1}^{d_i(<\infty)}\vert {i_k}\rangle
\langle {i_k}\vert$ and require that
$\sum_{ij}P_iP_j=\sum_iP_i=\hat 1$.  If all of the row sums
\begin{equation}
W_{ij,k}:=\hbox{$\sum_{k'=1}^{d_j}$}\,
h_{i_kj_{k'}},\quad k=1,\ldots, d_i,\label{eq:blocks}
\end{equation}
are independent of $k$, the Hamiltonian commutes with
the projection operator
$P:=\sum_i\vert \psi_i\rangle\langle\psi_i\vert$,
where $\vert \psi_i\rangle=d_i^{-1/2}
\sum_{k=1}^{d_i}\vert {i_k}\rangle$. Thus $H$ can be
written as a direct sum $H=H_{P}\oplus H_{\hat 1-P}$ with
matrix  elements of $H_P$ given as
\begin{equation}
h_{ij}:=\langle \psi_i\vert H\vert \psi_j\rangle
=W_{ij}(d_i/d_j)^{1/2}=W_{ji}^*(d_j/d_i)^{1/2}.
\label{eq:matrixel}
\end{equation}
The block-diagonalising transformation is not specific
to the present problem, as it amounts to extracting states 
with a specific symmetry from all of the system's 
eigenstates. The ground state of $H_{\rm C}+H_{\rm J}$ is always an
eigenstate of $H_P$, because $a_{\vec n}^{\vec q,\phi=0}>0$
for all $\vec q$ and $\vec n$. The Fourier expansion  (\ref{eq:amplit})
can be used simultaneously if each subspace label, $i$,
corresponds to a single value of  $Y_{\vec n}$. 

For homogeneous arrays at $\vec q=\vec n_0$
all junctions are indistinguishable in terms of the 
charging energy. The subspace labels 
are of the form  $\bar m_{\bar z}$ with  $d_{\bar m_{\bar z}}=N!/(
\prod_{j=1}^{j_{\rm max}}m_j!)$.
This subspace contains charge eigenstates carrying the label
\begin{equation}
\vec y=\left(z_1^{(1)},...,z_1^{(m_1)},z_2^{(1)},\ldots,z_2^{(m_2)},
\ldots,z_{j_{\rm max}}^{(m_{j_{\rm max}})}\right),
\end{equation}
where $z_1<\cdots<z_{j_{\rm max}}$ and
$m_1+\cdots+m_{j_{\rm max}}=N$, or
any distinct permutation of $\vec y$. Thus only the
number and multiplicity of tunnellings is of importance. 

On the gating path the gate charges
can be written as $\vec q=(1-x)\vec n_0+x\vec n_1$,
where $\vert\vec n_0\rangle$ ($\vert\vec n_1\rangle$) is the
initial (final) optimal charge state for the given leg.
Thus one junction becomes distinguishable and
subspace labels can be chosen as $z_0;\bar m_{\bar z}$, where
$\bar m_{\bar z}$ refers to the remaining $N-1$ components,
$0\le z_0+\sum_{i=1}^{N-1}m_iz_i<N$ and
$d_{z_0;\bar m_{\bar z}}=(N-1)!/(\prod_{j=1}^{j_{\rm max}}m_j!)$.

The diagonal and non-zero off-diagonal matrix elements 
of  $H_P$ are the  common charging energies and 
\begin{equation}
-(\varepsilon_{\rm J}m_j/2)
e^{\pm i\phi/N}(d_{(z_0;)\bar m_{\bar z}}/d_{(z'_0;)\bar m'_{\bar z'}})^{1/2}
\label{eq:matrixel2}
\end{equation}
with $m_0=1$ when applicable, respectively.
The block-diagonalised matrices are sparser than the 
original matrices and the amplitudes in eigenvectors are multiplied 
by $(d_{(z_0;)\bar m_{\bar z}})^{1/2}$ due to combining
of several amplitudes into one. A further block diagonalisation
is possible at $\vec q=\vec n_0$ and $\vec q=(\vec n_0+\vec n_1)/2$
if $\phi$ is a multiple of $\pi$.

A truncation picks the charge states required for 
reliable evaluation of eigenstates and observables.
First an initial truncation, a set of states
$B_i=\{\vert\vec n_j\rangle\}$, is chosen.
Its extensions, bases $B_i^{(l)}$, contain all neighbours
up to and including $l$th nearest neighbours of each and
every $\vert\vec n_j\rangle$.  For non-ideal cycles
and/or inhomogeneous arrays the initial truncation is the
``b basis'' of Ref.~\ref{bib:aun} which reproduces leading order
supercurrent and inaccuracy. A ``c basis'' is just the first order
extension of a ``b basis'', i.e. ``b basis$^{(1)}$''. 

For ideal cycles and homogeneous arrays
the optimal $B_0$ truncation  corresponds to $3N-2$ labels, that is
$0;(j,N-1-j)_{(0,1)}$, ``$1;(j,N-1-j)_{(-1,0)}$'', $j=0,\ldots,N-1$,
and $1;(j,N-1-j)_{(0,1)}$, $j=1,\ldots,N-2$. These
$3\cdot2^{N-1}-2$ charge states
actually contribute to the leading order supercurrent
and inaccuracy. More restricted, few-state
truncations $B_1=\{\vert\vec n_0\rangle\}$
and $B_2=\{\vert\vec n_0\rangle,\vert\vec n_1\rangle\}$ are
of use especially when the ground state energy is sought.
The efficiency of the block diagonalisation for bases $B_0^{(l)}$
is shown in Table~\ref{tab:dimen} where the number of labels
is compared against the  number of included charge eigenstates.
Examples of bases  $B_1^{(l)}$ and  $B_2^{(l)}$
are also given.

\vbox{
\begin{table}[hbt]
\caption{The number of subspace labels (sl) and included
charge states (ch) for  bases $B_0^{(l)}$ with several $N$
and examples of bases $B_1^{(l)}$ and $B_2^{(l)}$.
\label{tab:dimen}}
\begin{center}
\footnotesize
\begin{tabular}{ccccc}
{\raisebox{0pt}[3pt][6pt]{$N_{0/1/2}$}}&
sl$_{l}$&ch&sl$_{l}$&ch\cr
\hline
\raisebox{0pt}[10pt][1pt]{$4_0$}&$52_2$&$182$&
$1872_{14}$&$9572$\cr
\raisebox{0pt}[0pt][1pt]{$5_0$}&$82_2$&$626$&
$3082_{11}$&$46400$\cr
\raisebox{0pt}[0pt][1pt]{$6_0$}&$114_2$&$1918$&
$3468_{9}$&$1.56\cdot10^5$\cr
\raisebox{0pt}[0pt][1pt]{$7_0$}&$146_2$&$5428$&
$3941_{8}$&$5.22\cdot10^5$\cr
\raisebox{0pt}[0pt][1pt]{$8_0$}&$178_2$&$14498$&
$3648_{7}$&$1.30\cdot10^6$\cr
\raisebox{0pt}[0pt][1pt]{$9_0$}&$210_2$&$37082$&
$2899_{6}$&$2.44\cdot10^6$\cr
\raisebox{0pt}[0pt][1pt]{$8_1$}&$69_5$&$12331$&
$14727_{19}$&$6.15\cdot10^7$\cr
\raisebox{0pt}[0pt][1pt]{$7_2$}&$248_5$&$9452$&
$12842_{14}$&$2.19\cdot10^6$\cr
\end{tabular}
\end{center}
\end{table}
}

As a concrete example, take basis $B_1^{(l)}$ and the subspace labels 
$N_0$, $(N-1,1)_{(0,1)}$ and $(1,N-1)_{(0,1)}$,
standing for $2N+1$ charge eigenstates at $\vec q=\vec n_0$.
The block-diagonalised Hamiltonian obtained from Eqs.~(\ref{eq:matrixel})
and (\ref{eq:matrixel2}) is given by
\begin{equation}
H_{\rm bd}=\left(\matrix{0&K_N^*&K_N\cr
K_N&(N-1)/N&2\delta_{N3}K_N^*/\sqrt N\cr
K_N^*&2\delta_{N3}K_N/\sqrt N&(N-1)/N}\right)
\end{equation}
where $K_N:=-(\varepsilon_{\rm J}/2)e^{i\phi/N}\sqrt{N}$. For $N=3$
the leading order supercurrent for $\varepsilon_{\rm J}\ll1$ is
already reproduced. Note that labels $(N-1,1)_{(0,1)}$ and 
$(1,N-1)_{(0,1)}$ can be joined, if $\phi$ is a multiple of $\pi$.

The discussion has now lead to the main results of the paper, 
the general systematics of the pumped charge. For ideal gating
sequences $d{\cal M}=1/N$ for each leg due to symmetry.
Corresponding integrated pumped charge reads
\begin{equation}
Q_{\rm p}(N,\varepsilon_{\rm J},\phi,\vec c,{\rm leg})
:=1/N+\hbox{$\sum_{l=1}^{\infty}$}\,b_{l,{\rm leg}}\cos(l\phi),\label{eq:qpump}
\end{equation} 
which is always positive. Replace $1/N$ by $d{\cal M}$ for 
non-ideal cycles.
When performing numerical calculations,
the basis must contain all of the important charge states and
the number of angles used in fast Fourier transform, $2^{j}$,
must be large enough not to allow misidentification of  
non-negligible components $a_{\vec n,l}$ and $a_{\vec n,l+ 2^j}$ 
as a single component. Failure to satisfy
these requirements causes systematical error due to
incorrect amplitudes and spurious interference between
Fourier components in Eq.~(\ref{eq:amplit}), respectively.
The integration points should be chosen so that
the magnitude of norms $\Vert\,\vert dm\rangle\Vert$
is neither too large nor varies too much.

The rate of convergence for each $\cos(l'\phi)$-dependent term
in Eq.~(\ref{eq:qpump}) behaves as $1/(\#{\rm
steps})^2$. For a reasonable choice of integration points 300--500
steps per leg, much less than used in Refs.~\ref{bib:pek} and
\ref{bib:aun}, usually gives relative precision of the order of
$10^{-5}$ (disregarding the  systematical error) 
for coefficients with magnitude greater than $10^{-8}$.
This greatly diminishes convergence problems due to smallness of
$da_{\vec n,l}$ for large bases and large number of angles. 
The expression $2l'a_{\vec n,l}da_{\vec n,l+l'}$ has
been used as the latter part of Eq.~(\ref{eq:transchar}) for many of
the data points. In this case the rate of convergence behaves as $1/\#{\rm
steps}$. Bases $B_2^{(l)}$ have also been used when calculating 
$Q_{\rm p}$.

\vbox{
\begin{table}[hbt]
\caption{The first four ratios $\beta_l$ and the limiting value
$\tilde\beta$ for   $\varepsilon_{\rm J}=0.4$ as function
of $N$ and for $N=4$ as function of $\varepsilon_{\rm J}$.
Subscript of $\tilde\beta$ in the last column indicates 
the coefficient $\beta_l$ used when estimating the limiting value.
\label{tab:n4fourier}}
\begin{center}
\footnotesize
\begin{tabular}{cccccc}
{\raisebox{0pt}[3pt][4pt]{$(\varepsilon_{\rm J})_N$}}&$\beta_1$&
$\beta_2$&$\beta_3$&$\beta_4$&$(\tilde\beta)_l$\cr
\hline
\raisebox{0pt}[10pt][1pt]{$0.4_3$}&$0.8802$&$0.7782$&$0.7470$&$0.7347$
&$0.722_{60}$\cr
\raisebox{0pt}[8pt][1pt]{$0.4_4$}&$0.7250$&$0.5706$&$0.5354$&$0.5225$
&$0.509_{25}$\cr
\raisebox{0pt}[8pt][1pt]{$0.4_5$}&$0.5494$&$0.3842$&$0.3551$&$0.3451$
&$0.335_{20}$\cr
\raisebox{0pt}[8pt][1pt]{$0.4_6$}&$0.3844$&$0.2393$&$0.2188$&$0.2121$
&$0.206_{15}$\cr
\raisebox{0pt}[8pt][1pt]{$0.4_7$}&$0.2516$&$0.1401$&$0.1272$&$0.1231$
&$0.120_{11}$\cr
\raisebox{0pt}[8pt][1pt]{$0.4_8$}&$0.1563$&$0.0785$&$0.0708$&$0.0685$
&$0.067_{9}$\cr
\raisebox{0pt}[8pt][1pt]{$0.4_9$}&$0.0934$&$0.0427$&$0.0383$&$0.0370$
&$0.036_{7}$\cr
\raisebox{0pt}[8pt][1pt]{$0.4_{10}$}&$0.0542$&$0.0227$&$0.0203$&$0.0196$
&$0.019_{7}$\cr
\raisebox{0pt}[8pt][1pt]{$0.04_4$}&$0.0188$&$0.0096$&$0.0095$&$0.0095$&
$0.0095_4$\cr
\raisebox{0pt}[8pt][1pt]{$0.2_4$}&$0.3444$&$0.2164$&$0.2049$&$0.2025$&
$0.201_{14}$\cr
\raisebox{0pt}[0pt][1pt]{$0.6_{4}$}&$0.8853$&$0.7833$&$0.7469$&$0.7296$
&$0.701_{40}$\cr
\raisebox{0pt}[0pt][1pt]{$0.8_4$}&$0.9475$&$0.8876$&$0.8598$&$0.8444$
&$0.808_{80}$\cr
\raisebox{0pt}[0pt][1pt]{$1.0_4$}&$0.9738$&$0.9390$&$0.9198$&$0.9079$
&$0.870_{120}$\cr
\raisebox{0pt}[0pt][1pt]{$1.2_4$}&$0.9860$&$0.9654$&$0.9526$&$0.9441$
&$0.908_{160}$\cr
\end{tabular}
\end{center}
\end{table}
}

The sequence of the Fourier coefficients $\{b_{l,\rm leg}\}_{l=0}^\infty$, 
where $b_0=2/N$, is alternating and decreasing in magnitude.
Furthermore, the ratios $\beta_l:=\vert b_l/b_{l-1}\vert$, $l=1,2,\ldots$
form a decreasing sequence with limiting value
$\tilde\beta:=\lim_{l\rightarrow\infty}\beta_l
\ge b_0\beta_1$.\cite{note0} Because $\partial Q_{\rm p}/\partial \phi>0$
in the range $\phi\in(0,\pi)$, $Q_{\rm p}$ is bounded
from above by $[2(1-\beta_1)^{-1}-1]/N$ and from below by
$1/N-2\beta_1/(N+2\beta_1)$. For finite values of 
$\varepsilon_{\rm J}$ the ratios 
$\beta_l$ and $\tilde\beta/\beta_1$
are monotonously increasing functions 
of $\varepsilon_{\rm J}$ with limiting value of unity,
because $Q_{\rm p}(\phi=0,\varepsilon_{\rm J}
\rightarrow\infty)\searrow 0$, which enforces a stricter limit
for  the ratio $\tilde\beta/\beta_1$. This behaviour is explained
by increasing long-range (high-$l'$) correlations in 
Eq.~(\ref{eq:transchar}) or actually in the state itself.
On the other hand, the ratios are monotonously decreasing functions of $N$ 
for fixed $\varepsilon_{\rm J}$ as correlations are weakened in 
longer arrays. All of the above-mentioned features are clearly shown in
Table~\ref{tab:n4fourier}, where $\beta_l$ for the homogeneous case 
are depicted as function of $N$ at $\varepsilon_{\rm J}=0.4$
and as function of $\varepsilon_{\rm J}$ for $N=4$.\cite{note1}
These properties of the pumped charge are rather robust against 
relatively small systematical errors .

For small values of  $\varepsilon_{\rm J}\ll1$ 
one has $\beta_{l>1}\approx b_0\beta_1$ and 
$\beta_1\sim (N\varepsilon_{\rm J}/2)^{N-2}
[N(N-1)/2(N-2)!]$.
The effects due to inhomogeneity, that is the relative sizes of
$b_{1,{\rm leg}}$, can estimated by fixing the 
the leg index $r$ in the inhomogeneity prediction, 
Eq.~(37) of Ref.~\ref{bib:aun}. The corresponding results
work quite well, only slightly overestimating the ratio
between largest and and smallest coefficients, 
even for large inhomogeneities $X_{\rm inh}$.  
The total transferred charge $Q_{\rm p}
=\sum_{\rm leg} Q_{{\rm p},{\rm leg}}$ for inhomogeneous arrays
does not have to satisfy  $\beta_l>\beta_{l+1}$, 
although always $-b_{l+1}/b_l<1$. 
This symmetry is also broken by non-ideal gating sequences,
even for single legs. 

In order to conclude, the properties of the tunnelling-charging 
Hamiltonian of a Cooper pair pump have been examined using
an  efficient block-diagonalisation scheme and a
compatible Fourier expansion of the eigenstates. Explicit 
enforcement of  the model symmetries produces  strong systematics of 
the pumped charge, even if the structure of the Fourier coefficients
$\{b_{l,\rm leg}\}_{l=0}^\infty$ was not exhaustively proven.
These properties are possibly related to the
dynamical algebra of single and coupled Josephson junctions
described in Ref.~\ref{bib:cel}, which  offers a complementary
view of the present problem in case of a superconducting
loop.

\acknowledgements{This work has been supported by the Academy of Finland
under the Finnish Centre of Excellence Programme 2000-2005
(Project No. 44875, Nuclear and Condensed Matter Programme at JYFL).
The author thanks Mr. J.J. Toppari for insightful comments.
}



\begin{references}
\bibitem{pot92} H. Pothier, P. Lafarge, C. Urbina, D. Esteve
and M.H. Devoret, Europhys. Lett {\bf 17}, 249 (1992).
\bibitem{kel96} M. W. Keller, J. M. Martinis, N. M. Zimmerman and
A.H. Steinbach, Appl. Phys. Lett. {\bf 69}, 1804 (1996);
M.W. Keller, J.M. Martinis and R.L. Kautz, Phys. Rev. Lett.
{\bf 80}, 4530 (1998); M. W. Keller, A. L. Eichenberger, J. M. Martinis
and N. M. Zimmerman, Science {\bf 285}, 1716 (1999).
\bibitem{ave98}D. V. Averin, Solid State Commun. {\bf 105}, 659 (1998).
\bibitem{shn98}A. Shnirman and G. Sch\"on, Phys. Rev. B
{\bf 57}, 15400 (1998).
\bibitem{pek99} J. P. Pekola, J. J. Toppari, M. Aunola,
M. T. Savolainen and D. V. Averin, Phys. Rev. B {\bf 60}, 
R9931 (1999).\label{bib:pek}
\bibitem{aun00} M. Aunola, J. J. Toppari and J. P. Pekola, Phys. Rev. 
B {\bf 62}, 1296 (2000).\label{bib:aun}
\bibitem{pek00} J. P. Pekola and J. J. Toppari, 
submitted to Phys. Rev. Lett. (2000).\label{bib:top}
\bibitem{ing92} G.-L. Ingold and Yu. V. Nazarov,
in {\it Single Charge Tunnelling, Coulomb Blockade Phenomena
in Nanostructures}, eds.\ H.\ Grabert and M.L. Devoret,
(Plenum Press, New York, 1992).
\bibitem{rud73} W. Rudin, {\em Functional Analysis}, (McGraw Hill, 1973).
\bibitem{ber84} M. V. Berry, Proc. R. Soc. Lond.
A {\bf 392}, 45 (1984).
\bibitem{fal00}G. Falci, R. Fazio, G.M. Palma, J. Siewert,
and V. Vedral, Nature {\bf 407}, 355 (2000).\label{bib:fal}
\bibitem{note0}To some degree already in matrix elements 
$\langle m\vert\hat M\vert dm\rangle$.
\bibitem{note1}The full data set and some tools
for reproducing the data is available  at
http://www.jyu.fi/$\sim$mimaau/Cooper/ 
\bibitem{cel00} E. Celeghini, L. Faoro, and M. Rasetti,
Phys. Rev. B {\bf 62}, 3054 (2000).\label{bib:cel}
\end{references}
\end{document}